\documentclass[aps,prd,12pt,unsortedaddress,superscriptaddress,nofootinbib]{revtex4-1}
\usepackage{graphicx}
\usepackage{amsmath}
\usepackage{amsfonts}
\usepackage{amssymb,bm}
\usepackage{epstopdf}
\usepackage{hyperref}
\usepackage{float}
\usepackage[utf8]{inputenc}
\hypersetup{colorlinks = true,
           allcolors = blue}

\allowdisplaybreaks[1]
\begin{document}
\title{Triangle Singularities and Charmonium-like $XYZ$ States}
\date{\today}
\author{Feng-Kun Guo}
\affiliation{CAS Key Laboratory of Theoretical Physics, Institute of Theoretical Physics, Chinese Academy of Sciences, Beijing 100190, China}
\affiliation{School of Physical Sciences, University of Chinese Academy of Sciences, Beijing 100049, China}
\email{fkguo@itp.ac.cn}


\begin{abstract} 
 
The spectrum of hadrons is important for understanding the confinement of quantum chromodynamics. Many new puzzles arose since 2003 due to the abundance of experimental discoveries with the $XYZ$ structures in the heavy quarkonium mass region being the outstanding examples. 
Hadronic resonances correspond to poles of the $S$-matrix, which has other type of singularities such as the triangle singularity due to the simultaneous on-shellness of three intermediate particles. 
Here we briefly discuss a few possible manifestations of triangle singularities in the  $XYZ$ physics, paying particular attention to the formalism that can be used to analyze the data for charged $Z_c$ structures in the $\psi\pi$ distributions of the reaction $e^+e^-\to \psi\pi^+\pi^-$.
\end{abstract}
\maketitle

\section{Introduction}    

Being closely related to color confinement, hadron spectroscopy is one of the most important aspects of the nonperturbative quantum chromodynamics (QCD).
Many new hadron resonances or resonance-like structures have been observed since 2003 at worldwide experiments, including Belle, BaBar, BESIII, CDF, LHCb and so on. In particular, many of them were observed in the heavy-quarkonium mass region, and have properties difficult to be understood from the quark model point of view. Thus, they are called $XYZ$ states, and have spurred plenty of experimental and phenomenological investigations, as well as studies using lattice QCD. For recent reviews, we refer to Refs.\,[\citenum{Chen:2016qju,Esposito:2016noz,Hosaka:2016pey,Richard:2016eis,Lebed:2016hpi,Guo:2017jvc,Ali:2017jda,Olsen:2017bmm,Kou:2018nap,Cerri:2018ypt,Liu:2019zoy,Brambilla:2019esw,Guo:2019twa}].

\begin{figure}[tbh]
\begin{center}
\includegraphics[width=0.5\linewidth]{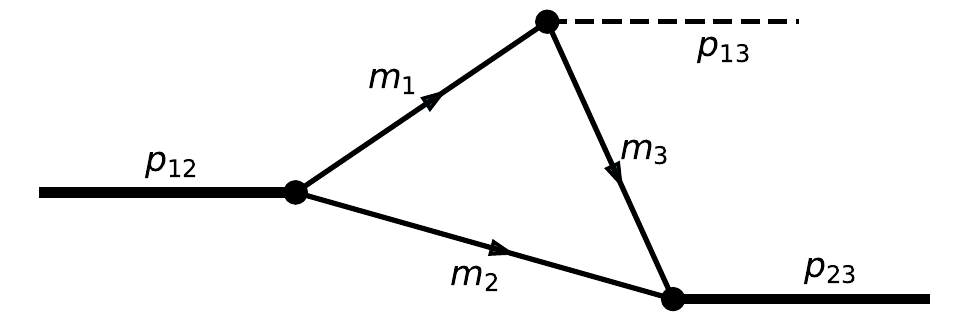}
\caption{A triangle diagram. Each external line does not necessarily represent a single particle.}
\label{fig:triangle}
\end{center}
\end{figure}

In order to understand the physics behind the messy spectrum of the $XYZ$ structures, we need to be careful about interpreting experimental observations. Most of these structures were discovered by observing a peaking structure in the invariant mass distribution of two or three particles in the final state. Peaking structures, in particular the narrow ones, are often due to singularities of the $S$-matrix, which have different kinds including poles and branch points. Resonances are poles of the $S$-matrix, while the branch points arise from unitarity and are due to the on-shellness of intermediate particles. 
The simplest one of the latter is the two-body threshold cusp. It is a square-root branch point and shows up exactly at all $S$-wave thresholds coupled to the measured energy distributions. The strength of the cusp depends on the masses of the involved particles and the interaction strength of the rescattering from the intermediate two particles to the final states.
Triangle singularity is more complicated. It is due to three on-shell intermediate particles, see Fig.~\ref{fig:triangle}, and happens on the physical boundary when the interactions at all the three vertices happen as classical processes in spacetime~\cite{Coleman:1965xm}. The triangle singularity is a logarithmic branch point, and thus can lead to drastic observable effects if it is located close to the physical region. 
The threshold cusp and triangle singularity are just two examples of the more general Landau singularities~\cite{Landau:1959fi}.
For a recent review of threshold cusps and TSs in hadronic reactions, we refer to Ref.\,[\citenum{Guo:2019twa}].

The triangle singularities related to the $XYZ$ structures have been discussed in Refs.~[\citenum{Wang:2013cya,Wang:2013hga,Liu:2013vfa,Liu:2014spa,Liu:2015cah,Szczepaniak:2015eza,Szczepaniak:2015hya,Albaladejo:2015lob,Liu:2016onn,Bondar:2016pox,Pilloni:2016obd,Gong:2016jzb,Guo:2019qcn,Braaten:2019gfj,Braaten:2019gwc,Braaten:2019yua,Braaten:2019sxh,Achasov:2019wvw,Nakamura:2019btl,Nakamura:2019emd,Nakamura:2019nwd}]. Here we focus on those related to the charged charmonium-like structures observed in the $e^+e^-\to J/\psi\pi^+\pi^-$ and $\psi(3686)\pi^+\pi^-$ reactions. 
Notice that the triangle singularity mechanism is not supposed to be the only reason behind the relevant observed peaks. 
In particular, there can be $Z_c$ resonances in addition, and the production of near-threshold resonances can get enhanced by the mechanism as emphasized in Ref.~[\citenum{Guo:2019twa}]. 
Thus, the formalism considering the triangle singularity effects should also have the freedom to include $Z_c$ contributions through the $\psi\pi$, $D\bar D^*$ and $D^*\bar D^*$ coupled-channel $T$-matrix.%
We shall present formulae that will be useful for an analysis of the data taking into account triangle singularities for such processes.  In Sec.~\ref{sec:2}, we discuss the relevant triangle diagrams for the reactions of interest, point out that both $S$- and $D$-wave couplings of the $D_1(2420)$ to the $D^*\pi$ need to be taken into account, and give the expressions which can be used to account for the triangle singularity effects in the analysis of the $e^+e^-\to \psi\pi^+\pi^-$ data. 
A brief summary is given in Sec.~\ref{sec:summary}. The scalar triangle loop integral is evaluated in the appendix.

\section{\texorpdfstring{Triangle diagrams for
\(e^+e^-\to \psi\pi^+\pi^-\) and the
amplitude}{Triangle diagrams for e\^{}+e\^{}-\textbackslash{}to \textbackslash{}psi'\textbackslash{}pi\^{}+\textbackslash{}pi\^{}- and the amplitude}}

\label{sec:2}

\subsection{Triangle singularity in the \texorpdfstring{$D_1\bar D D^*$}{D1DbarD*} diagram}

The location of the peak induced by a triangle singularity is normally not far from the corresponding two-body threshold. Thus, for the $Z_c(3900)$~\cite{Ablikim:2013mio,Liu:2013dau,Ablikim:2013xfr,Ablikim:2015swa,Collaboration:2017njt} and $Z_c(4020)$~\cite{Ablikim:2013emm,Ablikim:2013wzq}, it is important to discuss the triangle diagrams with intermediate $D\bar D^*+c.c.$
and $D^*\bar D^*$, respectively, coupled to the final states where the $Z_c$ structures were observed.
It has been pointed out in Refs.\,[\citenum{Wang:2013cya,Wang:2013hga,Albaladejo:2015lob,Gong:2016jzb}] that the $D_1\bar D D^*+c.c.$ loops are important for the understanding of the $Z_c(3900)$.\footnote{The broad $D_0^*(2400)$ was considered in Ref.\,[\citenum{Szczepaniak:2015eza}] instead of the narrow $D_1(2420)$.}

\begin{figure}[tbh]
\begin{center}
\includegraphics[width=0.5\linewidth]{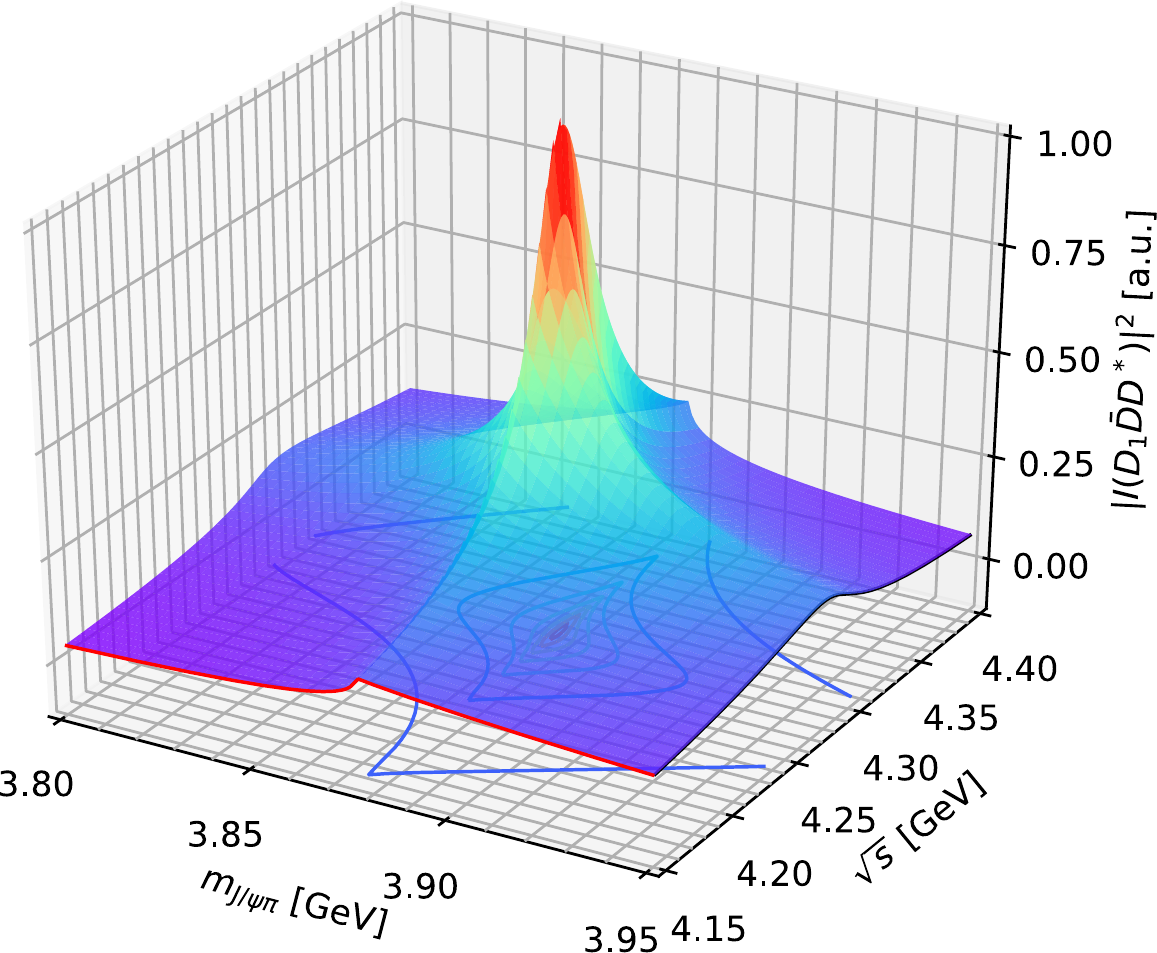}\\
\caption{The absolute value squared of the scalar three-point loop integral for the $D_1\bar D D^*$ intermediate particles as a function of the $\psi\pi$ invariant mass and the initial energy $\sqrt{s}$. }
\label{fig:3d}
\end{center}
\end{figure}

In order to see the possible impact of the $D_1\bar D D^*$ triangle diagram on the structures in both the final state $\psi\pi$ and initial state $\psi\pi\pi$ line shapes, a 3D plot for the $|I(D_1, \bar D, D^*)|^2$  is shown in Fig.~\ref{fig:3d}, where $I$ refers to the loop integral $I(m_1^2, m_2^2, m_3^2, p_{12}^2, p_{13}^2, p_{23}^2)$ defined in Eq.~\eqref{eq:Idef} in the appendix and we use the particle names to represent the corresponding $m_i^2$ and have neglected $p_{ij}^2$ for simplicity. The $\psi\pi$ pair comes from the $\bar D D^*$ rescattering (see Fig.~\ref{fig:triangle_ab}(b) below), and the $D_1$ width is taken into account by using a complex mass of the form $m_1-i\Gamma_1/2$. It is clear that the triangle loop integral is able to produce a peak in both the final state $\psi\pi$ invariant mass and the initial energy $\sqrt{s}$ distributions.

The sharp peak in the 3D plot is due to the presence of a triangle singularity which would be on the physical boundary if the width of the $D_1$ is neglected. One sees that the peak is around the $\bar D D^*$ threshold and the $D_1\bar D$ threshold in the $m_{J/\psi\pi}$ and $\sqrt{s}$ distributions, respectively. In addition, there is a clear cusp at the $\bar D D^*$ threshold in the $m_{J/\psi\pi}$ distribution, which is the manifestation of the two-body threshold as a subleading singularity of the triangle diagram. Because of the finite width of the $D_1$ there is no such an evident cusp in the $\sqrt{s}$ distribution. Because of the singular behavior, it is thus crucial to have the triangle diagrams included in the analysis of the data in order to extract the resonance parameters of the $Z_c$ or even to conclude whether it is necessary to introduce a $Z_c$. This is the point of Ref.\,[\citenum{Wang:2013cya}] which concluded the necessity of the $Z_c(3900)$, an opinion shared in Refs.\,[\citenum{Albaladejo:2015lob,Gong:2016jzb}], and suggested the importance of the $D_1\bar DD^*$ triangle singularity for the first time.

The BESIII data for both the $J/\psi\pi$ and $D\bar D^*$ invariant mass distributions were later on reanalyzed considering such triangle diagrams in Refs.\,[\citenum{Albaladejo:2015lob,Pilloni:2016obd}], while Ref.\,[\citenum{Albaladejo:2015lob}] concluded that there should exist a $Z_c(3900)$ as either a resonance pole above the $D\bar D^*$ threshold or a virtual state below it, Ref.\,[\citenum{Pilloni:2016obd}] concluded that the data could be fitted comparably well without introducing the $Z_c(3900)$.
One notable difference in the treatment of the triangle diagrams in these two references is that the $D_1D^*\pi$ coupling was treated as $D$-wave in Ref.\,[\citenum{Albaladejo:2015lob}] and as $S$-wave in Ref.\,[\citenum{Pilloni:2016obd}].

\subsection{The \texorpdfstring{$D_1D^*\pi$}{D1D*pi} coupling}
\label{sec:coupling}

The smallness of the $D_1(2420)$ width, $(31.7\pm2.5)$~MeV~\cite{Tanabashi:2018oca}, suggests that it is approximately a charmed meson with $j_\ell^P=\frac32^+$, where $j_\ell$ is the angular momentum of the light degrees of freedom in the $D_1$, including the light quark spin and the orbital angular momentum. Because $j_\ell^P=\frac12^-$ for the ground state $D^{(*)}$, a $\frac32^+$ meson decays into the $D^{(*)}\pi$ in a $D$ wave. Thus, the $D$-wave $D_1D^*\pi$ coupling was used in Refs.\,[\citenum{Wang:2013cya,Albaladejo:2015lob}].
However, it turns out that the $D$-wave decay can only account for about half of the $D_1(2420)$ decay width as we discuss now. There is another $\frac32^+$ charmed meson $D_2(2460)$ which is the spin partner of the $D_1(2420)$. Its decays into the $D^{(*)}\pi$ are purely $D$-wave. Thus, one can fix the $D$-wave decay coupling constant $h_D$ defined in the following Lagrangian (see Refs.\,[\citenum{Casalbuoni:1996pg,Colangelo:2005gb}] for the Lagrangian using the four-component notation)
\begin{equation}
  {\cal L}_D = \frac{h_D}{2F_\pi} {\rm Tr} \left[T_b^i\sigma^j H_a^\dag \right] \partial^i\partial^j\pi_{ba},
\end{equation}
which satisfies heavy quark spin symmetry (HQSS), where
\begin{align}
  H_a &= \vec{D}^*_a\cdot \vec{\sigma} + D_a, \nonumber\\
   T_a^i &= D_{2a}^{ij} \sigma^j + \sqrt{\frac23}\, D_{1a}^i + i \sqrt{\frac16}\, \epsilon_{ijk} D_{1a}^j \sigma^k
\end{align}
represent the $j_\ell=\frac12^-$ and $\frac32^+$ spin multiplets, respectively, $\vec\sigma$ is the Pauli matrices in the spinor space, Tr$[\cdot]$ denotes the trace in the spinor space, and $\pi_{ba}$ represents the pion fields with the subindices $a,b$ the indices in the light flavor space:
\begin{equation}
  \pi = \begin{pmatrix}
    \pi^0/\sqrt{2} & \pi^+ \\
    \pi^- & -\pi^0/\sqrt{2} 
  \end{pmatrix}.
\end{equation}
From reproducing the central value of the $D_2$ width, $(47.5\pm1.1)$~MeV, one gets $|h_D|=1.17~{\rm GeV}^{-1}$.
Using this value, one gets the $D$-wave contribution to the $D_1(2420)\to D^*\pi$ width as $15.2$~MeV, which is only about half of the $D_1(2420)$ width. Assuming that the $D^*\pi$ (and the sequential decay to $D\pi\pi$) modes dominate the $D_1$ width, the rest of the $D_1$ width, about 16.5~MeV, should come from $S$-wave decays. The $S$-wave $D_1D^*\pi$ coupling can be written as 
\begin{equation}
  {\cal L}_S = i\frac{h_S}{\sqrt{6}F_\pi} \vec D_{1b} \cdot \vec D^{*\dag}_a \partial^0 \pi_{ba} ,
\end{equation}
with the coupling constant $|h_S|=0.57$.
Here we require the $S$-wave coupling to be proportional to the pion energy to satisfy the Goldstone theorem because the pions are the pseudo-Goldstone bosons of the spontaneous breaking of chiral symmetry in QCD.

\subsection{\texorpdfstring{Amplitudes for \(e^+e^-\to \psi\pi^+\pi^-\) with \(THH\)-type triangle diagrams}{Amplitudes for e\^{}+e\^{}-\textbackslash{}to \textbackslash{}psi'\textbackslash{}pi\^{}+\textbackslash{}pi\^{}- with THH-type triangle diagrams}}

Let us now construct the amplitude of the triangle diagrams for the reaction \(e^+e^-\to \psi\pi^+\pi^-\), and we consider all the possible $THH$-type diagrams with $T$ and $H$ representing the $j_\ell^P = \frac32^+$ and $\frac12^-$ charmed mesons, respectively.

\begin{figure}
\begin{center}
\includegraphics[width=0.45\linewidth]{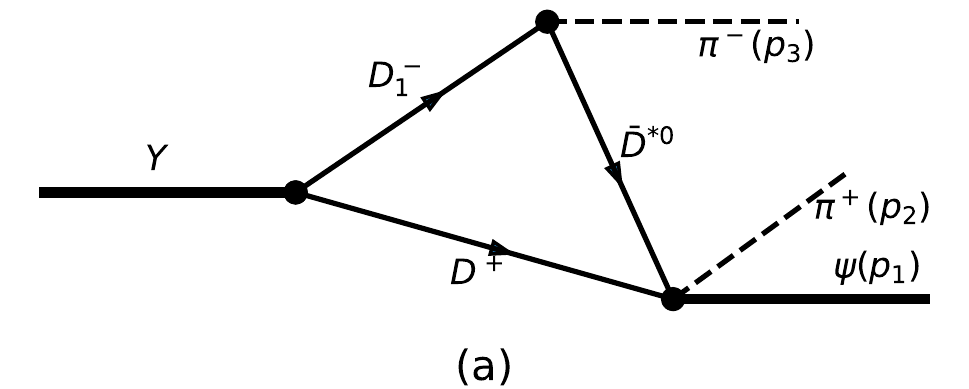}\qquad
\includegraphics[width=0.45\linewidth]{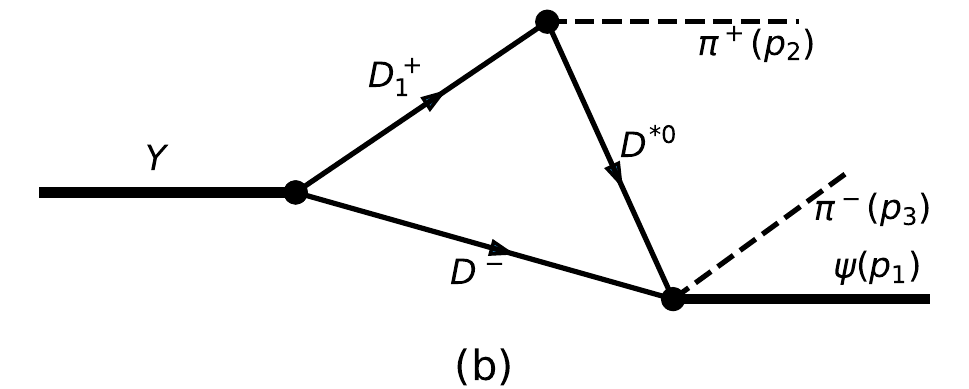}\\
\caption{The $D_1\bar D D^*+c.c.$ triangle diagrams with charged $D_1\bar D$  contributing to the process $Y\to \psi\pi^+\pi^-$ with the initial $Y$ coming from $e^+e^-$. }
\label{fig:triangle_ab}
\end{center} 
\end{figure}

The relevant $THH$-type triangle diagrams include  $D_1\bar D D^*+c.c.$,  $D_1\bar D^* D^*+c.c.$, $D_2\bar D^* D+c.c.$ and $D_2\bar D^* D^*+c.c.$ The $D_1\bar D D^*+c.c.$ diagrams for the charged $D_1\bar D$ are shown in Fig.~\ref{fig:triangle_ab}, and the analogous diagrams for the neutral $D_1\bar D$ are not shown. The diagrams for the other mentioned $THH$ triangles are similar. They were considered in Ref.\,[\citenum{Liu:2014spa}].
Here we only consider the narrow $D_{1(2)}$ mesons. However, one should keep in mind that their production together with a $\bar D^{(*)}$ is suppressed in the heavy quark limit~\cite{Li:2013yka}, and the broad $D_{0}(D_1')$ might also play a role here. But the $D_{0}(D_1')$ properties are still under discussion and the values listed in the Review of Particle Physics (RPP)\cite{Tanabashi:2018oca}, where were extracted from fitting to the $D^{(*)}\pi$ invariant mass distributions using the Breit-Wigner parametrization, are not trustworthy (for detailed discussion, see Refs.\,[\citenum{Du:2017zvv,Du:2019oki}]).

For the narrow $D_1(2420)$ and $D_2(2460)$ decays into $D^{(*)}\pi$, we use the coupling discussed in Sec.~\ref{sec:coupling}.
In principal, the rescattering from the  intermediate $D^{(*)}\bar D^{(*)}$ into $\psi\pi$ needs to be described through a coupled-channel $T$-matrix, see the treatment in Ref.\,\cite{Albaladejo:2015lob}. For simplicity, one may also approximate the $T$-matrix by that from a $Z_c$ exchange with the $Z_c$ parametrized using a Flatt\'e form to account for the $D^{(*)}\bar D^*$ threshold.

With the $Y, Z_c$ and $\pi$ as the external particles, the amplitude for a triangle loop mentioned above shown as Fig.~\ref{fig:triangle_ab}(a) with a $D$-wave $TH\pi$ coupling is proportional to 
\begin{align}
 {\cal A}_{D,a} \propto & I(m_1^2, m_2^2, m_3^2, s, M_\pi^2, p_Z^2)  \left(3 \vec{p}_3\cdot \vec\varepsilon_{Y} \vec{p}_3\cdot \vec\varepsilon_{Z} - \vec{p}_3^{\,2} \,\vec\varepsilon_{Y}\cdot \vec\varepsilon_{Z} \right), 
 \label{eq:AD}
\end{align}
and that with an $S$-wave $D_1D^*\pi$ coupling is proportional to
\begin{equation}
   {\cal A}_{S,a} \propto I(m_1^2, m_2^2, m_3^2, s, M_\pi^2, p_Z^2) E_3\vec\varepsilon_{Y}\cdot \vec\varepsilon_{Z} ,
   \label{eq:AS}
\end{equation}
where $s$ is the c.m. energy squared for the $e^+e^-$, $p_Z^2$ is the invariant mass squared of the meson pair coupled to the $Z_c$, and $\vec\varepsilon_{Y(Z)}$ are the spatial components of the polarization vectors for the $Y(Z_c)$. The polarization sum for each of them is $\sum_{\lambda}\varepsilon_{(\lambda)}^i \varepsilon_{(\lambda)}^j = \delta^{ij}$.
Notice that the intermediate charmed mesons are treated nonrelativistically, so that the Lorentz boost effect from the $D_1$ rest frame to the $e^+e^-$ c.m. frame is of higher order. Thus, the partial waves in the $D_1D^*\pi$ coupling lend directly to the partial waves between the pion and $Z_c$ (or the pair of the $\psi$ and the other pion).  
The intermediate particles in the diagram shown in Fig.~\ref{fig:triangle_ab}(b) is charge conjugated to those in the one in Fig.~\ref{fig:triangle_ab}(a). The amplitude can be obtained by changing $p_3$ to $p_2$ in the above expressions. 

Here let us give expressions and relations for some kinematic variables entering into the analysis. We define the following variables:
\begin{align}
  m_{12}^2 = (p_1+p_2)^2, \quad  m_{13}^2 = (p_1+p_3)^2, \quad  m_{23}^2 = (p_2+p_3)^2 .
\end{align}
They satisfy
\begin{align}  
m_{12}^2 + m_{13}^2 + m_{23}^2 = s + \sum_{i=1,2,3} M_i^2,
\end{align}
where $M_i$'s are the masses of the external particles in the final state. In terms of these variables, $p_Z^2$ in Eqs.~\eqref{eq:AD} and \eqref{eq:AS} is $m_{12}^2$, and that for Fig.~\ref{fig:triangle_ab}(b) is $m_{13}^2$.

In the rest frame of the initial state, we express all momenta and energies in terms of $m_{12}$ and $m_{23}$:
\begin{align}
  |\vec{p}_2| &= \frac1{2 \sqrt{s}} \sqrt{\lambda(s, M_2^2, m_{13}^2)}, \quad
  |\vec{p}_3| = \frac1{2 \sqrt{s}} \sqrt{\lambda(s, M_3^2, m_{12}^2)}, \nonumber\\
  E_2 &= \sqrt{M_2^2 + \vec{p}_2^{\;2}}, \qquad\qquad\quad~
  E_3 = \frac1{2\sqrt{s}}\left( s+M_3^2-m_{12}^2\right), \nonumber\\
  \vec{p}_2\cdot \vec{p}_3 & = \frac12 \left(M_2^2+M_3^2 + 2E_2E_3 -m_{23}^2 \right).
\end{align}

\begin{figure}[tbh]
\begin{center}
\includegraphics[width=0.45\linewidth]{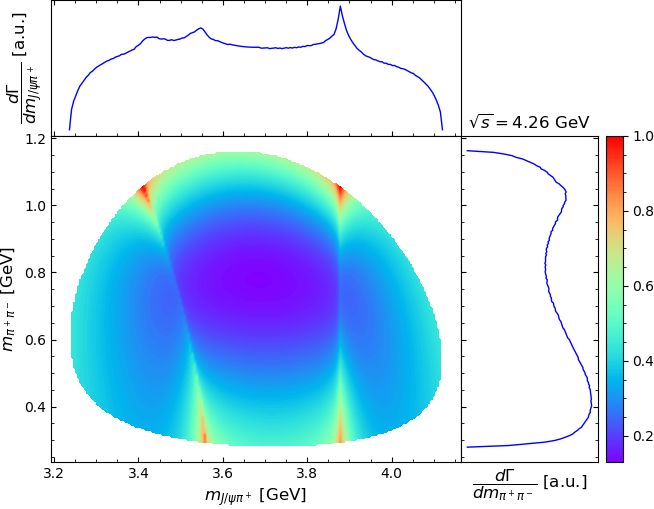}\qquad
\includegraphics[width=0.45\linewidth]{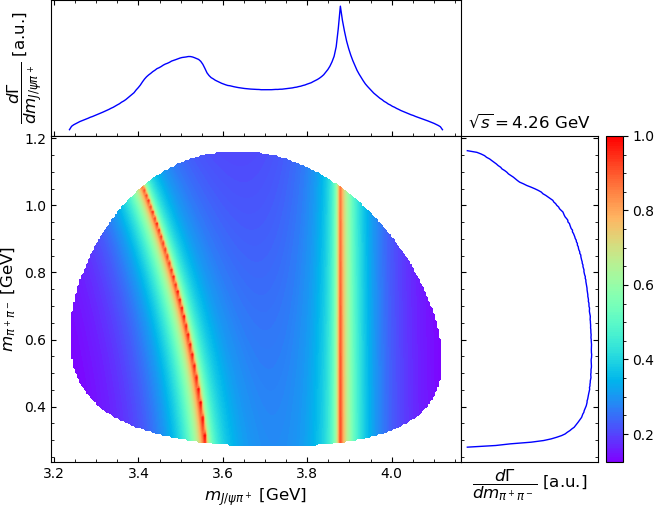}\\[2mm]
\includegraphics[width=0.45\linewidth]{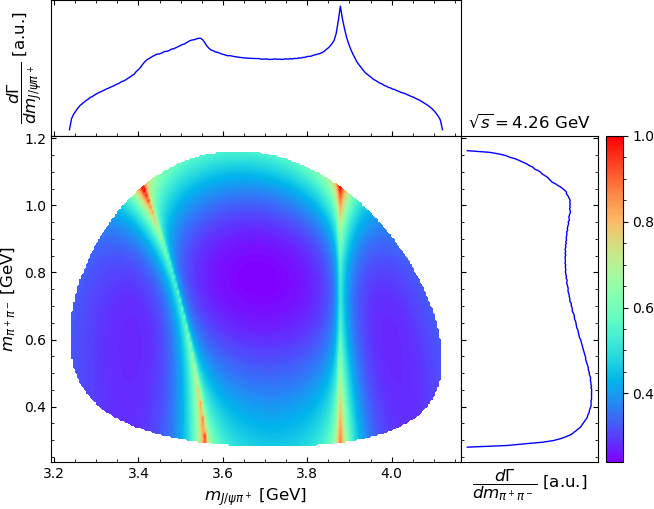}\hfill\\
\caption{ The Dalitz plot and its projections to the $J/\psi\pi^+$ and $\pi^+\pi^-$ energy distributions for the $e^+e^-\to J/\psi\pi^+\pi^-$ reaction induced by the $D_1\bar D D^*$ triangle diagrams. The $e^+e^-$ c.m. energy is taken at 4.26~GeV. The first row: $D$-wave $D_1D^*\pi$ coupling (left) and $S$-wave $D_1D^*\pi$ coupling (right); the second row: the coupling contains both the $S$-wave and $D$-wave parts as given in Sec.~\ref{sec:coupling}.
}
\label{fig:D1DDstar_SD}
\end{center}
\end{figure}

Before we move on, let us first show the impact of considering different partial waves for the $D_1D^*\pi$ coupling in the structure induced by the $D_1\bar D D^*+c.c.$ triangle diagrams for the $e^+e^-\to J/\psi\pi^+\pi^-$ reaction in Fig.~\ref{fig:D1DDstar_SD}. 
The sharp peak in the $J/\psi\pi^+$ energy distribution (the right band in the Dalitz plot) is due to the diagram in Fig.~\ref{fig:triangle_ab}(a), and the broader peak (the left band in the Dalitz plot) is the kinematical reflection due to the diagram in Fig.~\ref{fig:triangle_ab}(b).
One sees that the pion-momentum dependence at the $D_1D^*\pi$ vertex has a significant impact on the distributions. 
The reason is that the magnitude of the three-momentum of a pion can be as large as about 1~GeV, and a larger background in the $J/\psi\pi^+$ distribution can be caused by the $D$-wave coupling than that by the $S$-wave one. 
This could be the main reason for the different conclusions reached in Refs.\,[\citenum{Albaladejo:2015lob,Pilloni:2016obd}]. One also sees that the $D$-wave coupling also leads to a double-bump structure in the $\pi^+\pi^-$ invariant mass distribution.

Next, let us consider all the $THH$-type diagrams.
For accounting for the whole set of the $THH$-type triangle diagrams, 
one needs to decide on the relative couplings between $D_1\bar D+c.c.$, $D_1D^*+c.c.$ and $D_2\bar D^*+c.c.$ with the initial state.
In principal, for an analysis of the experimental data, one may assume them to be independent. This is because at different $e^+e^-$ c.m. energies, the charmonium or charmonium-like state that is important in that energy region could be different internal structures. This makes difficult the use of HQSS to relate these couplings.
Then, we can write the amplitude for Fig.~\ref{fig:triangle_ab}(a) considering all the $THH$-type diagrams as:
\begin{align}
 {\cal A}_a = &\, i \Bigg\{ 10h_S\,  E_3 \vec\varepsilon_Y\cdot
   \vec\varepsilon_\psi \left[ c_1 I\left(D_1,\bar{D},D^{\text{*}},s,M_3^2, m_{12}^2\right)T_*(m_{12}^2) \right.\nonumber\\
   &\, \left. +c_2 I\left(D_1,\bar{D}^{\text{*}},D^{\text{*}}, s,M_3^2, m_{12}^2\right) T_{**}(m_{12}^2)\right]   + h \left(3 \vec{p}_3\cdot \varepsilon_Y \vec{p}_3\cdot \vec\varepsilon_\psi-\vec{p}_3^{\;2} \vec\varepsilon_Y\cdot \vec\varepsilon_\psi\right) \nonumber\\
 &\, \times\left[10 c_1 I\left(D_1,\bar{D},D^{\text{*}},s,M_3^2, m_{12}^2\right)T_*(m_{12}^2) \right.  -5 c_2
   I\left(D_1,\bar{D}^{\text{*}},D^{\text{*}},s,M_3^2, m_{12}^2\right)T_{**}(m_{12}^2)  \nonumber\\
 &\, \left. -3 c_3 I\left(D_2,\bar{D}^{\text{*}},D^{\text{*}},s,M_3^2, m_{12}^2\right)T_{**}(m_{12}^2) -2 c_3
   I\left(D_2,\bar{D}^{\text{*}},D,s,M_3^2, m_{12}^2\right)T_{*}(m_{12}^2)\right]  \Bigg\}\nonumber\\
   \equiv & B_{S,a} \vec\varepsilon_Y\cdot \vec\varepsilon_\psi  + B_{D,a} \left(3 \vec{p}_3\cdot \vec\varepsilon_Y \vec{p}_3\cdot \vec\varepsilon_\psi-\vec{p}_3^{\;2} \vec\varepsilon_Y\cdot \vec\varepsilon_\psi\right) ,
   \label{eq:Aa}
\end{align}
with $M_3=M_2=M_\pi$. Here, we use $c_{1,2}$ and $c_3$ to account for the couplings of the $D_1\bar D$, $D_1 D^*$ and $D_2\bar D^*$ pairs to the initial state, respectively, and use the intermediate particles to represent the corresponding $m_i^2$.
$T_*$ and $T_{**}$ represent the $T$-matrix elements (with the polarization vectors amputated) for the rescattering processes $\bar D D^*\to\psi\pi^+$ and $\bar D^* D^*\to\psi\pi^+$, respectively.
The amplitude for the diagrams with charge-conjugated intermediate particles, ${\cal A}_b$, is obtained by replacing $M_3$ by $M_2$, $\vec p_3$ by $\vec p_2$, $E_3$ by $E_2$, and $m_{12}$ by $m_{13}$:
\begin{equation}
  {\cal A}_b = B_{S,b} \vec\varepsilon_Y\cdot \vec\varepsilon_\psi  + B_{D,b} \left(3 \vec{p}_2\cdot \vec\varepsilon_Y \vec{p}_2\cdot \vec\varepsilon_\psi-\vec{p}_2^{\;2} \vec\varepsilon_Y\cdot \vec\varepsilon_\psi\right).
  \label{eq:Ab}
\end{equation}
The sum ${\cal A}_a + {\cal A}_b$, after having parameterized $T_*$ and $T_{**}$ for the $D^{(*)}\bar D^{(*)}\to \psi\pi$ rescattering which allows for the existence of a $Z_c$ pole, may be used to account for the triangle singularity effects in the analysis of the data for the $e^+e^-\to \psi\pi^+\pi^-$ reaction.

In the following, we choose specific ratios among $c_{1,2,3}$ to show the triangle-singularity induced structures. Here, we assume that the initial vector source couples to the $T\bar{H}+c.c.$ pairs like a $D$-wave ($j_\ell=2$) charmonium. That is to assume that the HQSS breaking happens at the charmonium level, instead of at the charmed meson level. 
The reason for this choice is that the $D_1$ width is small so that the $D_{1(2)}$ mesons are well approximated by $j_\ell^P=\frac32^+$ mesons and the $S$-$D$ mixing for the $D_1$ mesons should be small (the mixing angle was determined to be $-0.10\pm0.04$ rad by Belle~\cite{Abe:2003zm}, noting however that the broad $D_1$ extracted from that reference is problematic as discussed in Refs.\,[\citenum{Du:2017zvv,Du:2019oki}]), while higher charmonia typically have a large $S$-$D$ mixing as can be seen from the fact that their dileptonic decay widths do not differ much~\cite{Tanabashi:2018oca} (see the discussion in Ref.\,\cite{Wang:2013kra}; the $S$-$D$ mixing angle for vector charmonia above 4~GeV could be as large as about 34$^\circ = 0.59$\,rad\,\cite{Badalian:2008dv}).
This assumption amounts to take $c_1=c_2=c_3$ in Eq.~\eqref{eq:Aa}.
Then, the absolute value squared of the amplitude reads as
\begin{align}
 |{\cal A}_a +{\cal A}_b|^2 =&\, 6 \Bigg\{ \frac12\left(|B_{S,a}|^2 + |B_{S,b}|^2\right) + {\rm Re} \left(B_{S,a} B_{S,b}^*\right) \nonumber\\
 &\,+ |B_{D,a}|^2\vec p_3^{\;4} 
 + |B_{D,b}|^2\vec p_2^{\;4} +  {\rm Re} \left(B_{D,a} B_{D,b}^*\right)  \left[ 3(\vec p_2\cdot \vec p_3)^2 - \vec p_2^{\;2} \vec p_3^{\;2}\right]
 \Bigg\}.
\end{align}

\begin{figure}[tbh]
\begin{center}
\includegraphics[width=0.48\linewidth]{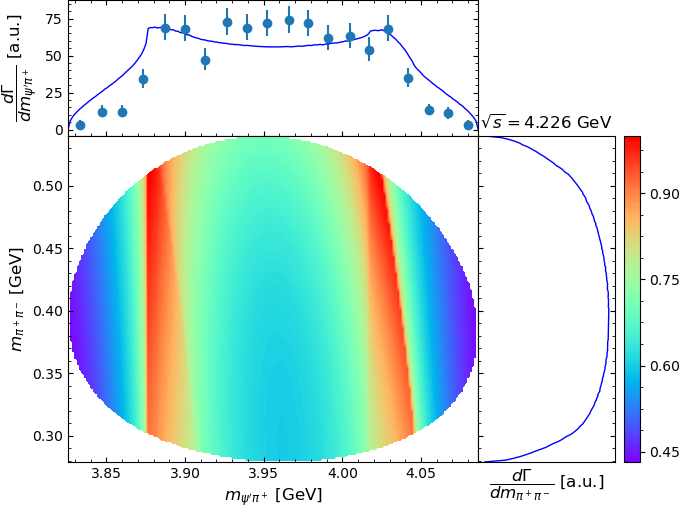}\hfill
\includegraphics[width=0.48\linewidth]{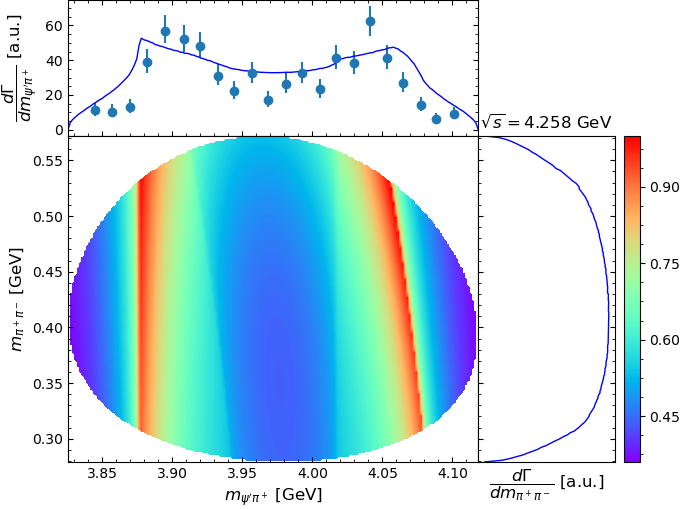}\\[2mm]
\includegraphics[width=0.48\linewidth]{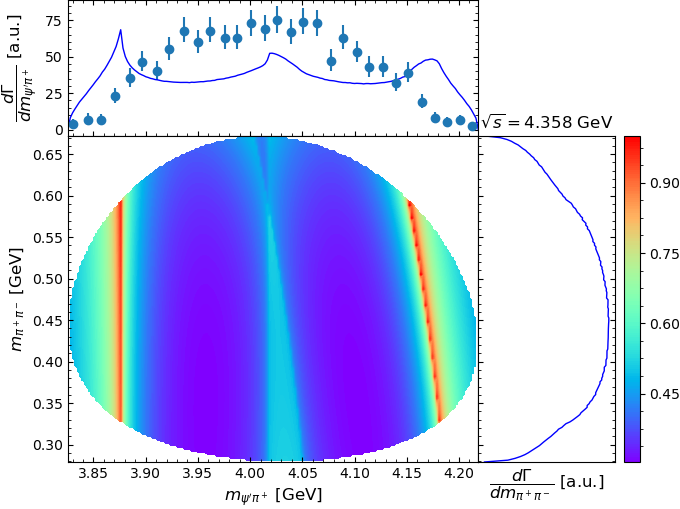}\hfill
\includegraphics[width=0.48\linewidth]{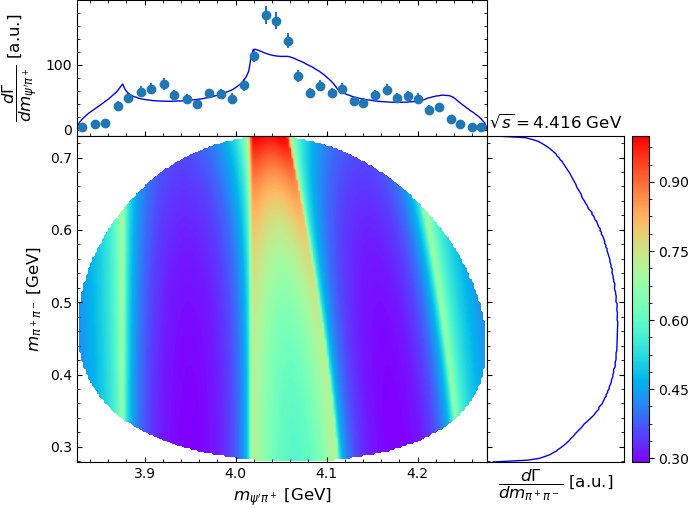}
\caption{The Dalitz plot and its projections to the $\psi'\pi^+$ and $\pi^+\pi^-$ energy distributions for the $e^+e^-\to \psi'\pi^+\pi^-$ reaction induced by the $THH$-type triangle diagrams. The amplitude is given by the sum of Eqs.~\eqref{eq:Aa} and \eqref{eq:Ab} with  $T_*$ and $T_{**}$ set to the same constant and $c_1=c_2=c_3$.
The BESIII data of the $\psi(3686)\pi$ invariant mass distributions~\cite{Ablikim:2017oaf} are shown for a rough comparison.
 }
\label{fig:THH}
\end{center}
\end{figure}

In order to see clearly the triangle singularity effects, we also switch off any nontrivial structure in the rescattering matrix elements $T_*$ and $T_{**}$ by setting them to the same constant.
The resulting Dalitz plot distributions for $e^+e^-\to \psi(3686)\pi^+\pi^-$ at four different c.m. energy values, $\sqrt{s}=4.226$, 4.258, 4.358 and 4.416~GeV, are shown in Fig.~\ref{fig:THH}. 
We also show the BESIII data of the $\psi(3686)\pi$ invariant mass distributions reported in Ref.~[\citenum{Ablikim:2017oaf}] for a rough comparison. 
Notice that we did not perform a fit to the data. 
The sensitivity of the triangle-singularity induced structure on the energy is evident, as already observed for this reaction in Ref.~[\citenum{Liu:2014spa}]. 
For a fit to the data, which is beyond the scope of this proceedings, one needs to parameterize the rescattering $T$-matrix, which contains possible $Z_c$ states, and treat $c_{1,2,3}$ as free parameters for each energy point. In addition, one also needs to include contributions other than the triangle diagrams such as the direct production of $\psi(3686)\pi^+\pi^-$ and the two-body rescattering (including the $\pi\pi$ final state interaction and the $T_*$ and $T_{**}$ as those in Eq.~\eqref{eq:Aa}).%
The BESIII data for this reaction reported in Ref.~[\citenum{Ablikim:2017oaf}] was analyzed in Ref.~[\citenum{Molnar:2019uos}] without considering the triangle singularities.

\section{Summary}
\label{sec:summary}

The triangle singularity effects need to be properly taken into account in order to establish the exotic hadron spectrum and extract the resonance parameters more reliably.
In this paper, we present the formalism of considering the triangle singularities that can be used in the experimental analysis of the $e^+e^-\to \psi\pi^+\pi^-$ data.
Notice that in a complete analysis, 
the triangle diagrams are not supposed to provide the only mechanism for the reactions. For instance, there can be a direct production of the $J/\psi\pi\pi$ and $D^{(*)}\bar D^*\pi$, followed by $\pi\pi$ and $D^{(*)}\bar D^*$-$J/\psi\pi$ final state interactions. While the $D^{(*)}\bar D^*$-$J/\psi\pi$ final state interactions are the same as $T_*$ and $T_{**}$ in Eq.~\eqref{eq:Aa},%
the $\pi\pi$ final state interaction also needs to be taken into account, which may be done by using the Omn\`es dispersive formalism as that used in Refs.~[\citenum{Chen:2019mgp,Molnar:2019uos}].

\bigskip

\section*{Appendix: Calculating the triangle loop integral}
\label{sec:app}

The essential function for evaluating amplitudes with triangle singularities is the following three-point scalar one-loop integral (for definitions of masses and the external momenta, see Fig.~\ref{fig:triangle}):
\begin{align}
  &I(m_1^2, m_2^2, m_3^2, p_{12}^2, p_{13}^2, p_{23}^2) \nonumber\\ 
  =&\, i \int \frac{d^4q}{(2\pi)^4} \frac{1} { (q^2-m_1^2 + i\epsilon) [ (p_{12}-q)^2 -m_2^2 + i\epsilon ][ (q-p_{13})^2-m_3^2+i\epsilon ]} 
  \label{eq:Idef}\\
  \equiv &\,  i \int \frac{d^4q}{(2\pi)^4} \frac{1} { J_1 J_2 J_3 } . \nonumber
\end{align}
This loop integral is ultraviolet convergent.
Using the method of Feynman parameters for this integral, we have
\begin{align}
  &I(m_1^2, m_2^2, m_3^2, p_{12}^2, p_{13}^2, p_{23}^2) \nonumber \\
  =&\, i \int  \frac{d^4q}{(2\pi)^4} \int d\alpha_1  d\alpha_2 \frac{2}{\left[\alpha_1 J_1 + \alpha_2 J_2 + (1-\alpha_1-\alpha_2)J_3\right]^3} \nonumber\\
  =&\, i \int  \frac{d^4q}{(2\pi)^4} \int_0^1 d\alpha_1 \int_0^{1-\alpha_1} d\alpha_2 \frac{2}{\left(q^2 -\Delta \right)^3} \nonumber\\
  =&\, \frac{1}{4\pi^2} \int_0^1 d\alpha_1 \int_0^{1-\alpha_1} d\alpha_2 \frac1{\Delta},
  \label{eq:int}
\end{align}
where 
\begin{align}
  \Delta &= p_{23}^2 \left(\alpha_2^2 - b\alpha_2 +c \right) , \nonumber\\
  b & = 1 + \frac1{p_{23}^2} \left[ \alpha_1(p_{12}^2 - p_{13}^2-p_{23}^2) + m_3^2 - m_2^2 \right], \nonumber\\
  c & = \frac1{p_{23}^2} \left[m_3^2 + \alpha_1(m_1^2-m_3^2) - \alpha_1(1-\alpha_1) p_{13}^2 \right] - i\epsilon. \nonumber
\end{align}
The integral over $\alpha_2$ in Eq.(\ref{eq:int}) can be worked out. For $4c>b^2$, one gets
\begin{align}
  &I(m_1^2, m_2^2, m_3^2, p_{12}^2, p_{13}^2, p_{23}^2) \nonumber\\
  =&\,
  \frac1{8\pi^2 p_{23}^2} \int_0^1d\alpha_1 \frac1{\sqrt{4c-b^2}} \Bigg[\arctan \frac{b}{\sqrt{4c-b^2}}  - \arctan \frac{b+2(\alpha_1-1)}{\sqrt{4c-b^2}} \Bigg].
  \label{eq:Ieval}
\end{align}
The remaining integration can be easily computed numerically.

\section*{Acknowledgements}

This work reported is supported in part by the National Natural Science Foundation of China (NSFC) and  the Deutsche Forschungsgemeinschaft (DFG) through the funds provided to the Sino-German Collaborative Research Center  CRC110 ``Symmetries and the Emergence of Structure in QCD"  (NSFC Grant No. 11621131001), by the NSFC under Grants No. 11835015, No.~11947302 and No. 11961141012, by the Chinese Academy of Sciences (CAS) under Grants No. QYZDB-SSW-SYS013 and No. XDB34030303, and by the CAS Center for Excellence in Particle Physics (CCEPP).


\end{document}